\documentclass[aps,prl,reprint,superscriptaddress]{revtex4-1}

\usepackage{amsmath,amssymb,bm,physics,graphicx}

\begin{document}

\title{Contextual Light-Particle Interference}

\author{Brian Stout}
\affiliation{Aix-Marseille Univ, CNRS, Centrale Marseille, Institut Fresnel, 13013 Marseille, France}

\begin{abstract}
We propose and analyze a generalized model of photodetection in which the detector couples simultaneously to both electric and magnetic components of the electromagnetic field. Extending Glauber's seminal theory we show that such detectors, now experimentally feasible via meta-materials, can yield interference patterns that depend strongly on the internal response of the detector. We show that for a two-source interference setup, coherent electric-magnetic dipole coupling can suppress or enhance interference visibility, revealing a clear instance of measurement contextuality in quantum optics, where the outcome depends on how the system is probed.
\end{abstract}

\maketitle

\textit{Introduction.}~Glauber’s quantum theory of optical coherence \cite{Glauber1963} is a cornerstone of modern quantum optics. In this formalism, the primary photodetection probability is the normally ordered correlation function: \begin{align}
P = s^{2}\langle \widehat{E}^{(-)}(\mathbf{r},t)\widehat{E}^{(+)}(\mathbf{r},t) \rangle \;, \label{eq:GlauberRule}
\end{align} which underlies the standard definition of intensity and remains central in both theoretical and experimental treatments of quantum light \cite{Loudon,Bachor,BarnettRadmore}. The parameter, $s$, is a measured empirical \emph{sensitivity} of the detector that will be important for subsequent discussions.

This framework has successfully explained a wide range of quantum optical phenomena, including photon bunching, antibunching, and two-photon interference effects \cite{MandelWolf,Aspect1981,Hong1987}. Nevertheless, improvements in single-photon detection have led to the development of meta-material and superconducting technologies that now enable detectors responsive to both electric and magnetic fields in their efforts to approach 100$\%$ quantum detector efficiencies, which goes well beyond the electric-dipole interactions that underlie Glauber's theory. 

Furthermore, recent theoretical developments, motivated by ideas from atomic superradiance and subradiance \cite{Dicke1954,Gross1982,Scully2006}, have raised questions about the completeness of the Glauber formalism in scenarios involving structured quantum interference.\cite{Villas-Boas2025} We develop a generalized theory of photodetection that includes coherent contributions from both electric and magnetic dipole interactions. In this extended framework, the usual intensity operator is replaced by a bilinear observable combining coherent contributions from the $\widehat{\boldsymbol{E}}$ and $\widehat{\boldsymbol{B}}$ field operators. This modification opens the door to interference behaviors that are strictly absent in the traditional Glauber theory, and invites an alternative interpretive framework for quantum optics where interference arises not as an intrinsic field quantity, but rather as a contextual phenomenon where it is induced by the manner in which light interacts with matter.

In this Letter, we explore these ideas in the context of a minimal quantum model involving two spatially separated dipole sources. We show that this system can exhibit context-dependent interference effects, which deviate measurably from the predictions of standard photodetection theory. While rooted in a familiar classical picture of radiated field interference, our treatment is fully quantum and extends the conceptual reach of Glauber's original framework. We argue that this approach provides a natural foundation for describing a broader class of quantum optical interference phenomena, especially in systems with complex emitter-detector couplings.

\textit{Generalized Detection Formalism.}~We consider a photodetector whose interaction Hamiltonian is of the form:
\begin{align}
\widehat{H}_{\mathrm{int}} \propto \mathbf{u}_{e} \cdot \widehat{\boldsymbol{E}}(\mathbf{r},t) + \mathbf{u}_{b} \cdot \widehat{\boldsymbol{B}}(\mathbf{r},t),
\end{align}
where $\mathbf{u}_{e}$ and $\mathbf{u}_{b}$ are unit vectors that describe the detector's internal electric and magnetic dipole sensitivity vectors. 
The generalized detection operator is defined as:
\begin{align}
\widehat{\mathcal{O}} = \mathbf{u}_{e}^{*} \cdot \widehat{\boldsymbol{E}}^{(+)}(\mathbf{r},t) + \zeta \, \mathbf{u}_{b}^{*} \cdot \widehat{\boldsymbol{B}}^{(+)}(\mathbf{r},t) \;, \label{Genop}
\end{align}
where the dimensionless mixing parameter, $\zeta$, may be complex in general, thus allowing for a relative weighting in both amplitude and phase of the electric and magnetic field contributions. The detection probability becomes $P \propto \langle \widehat{\mathcal{O}}^{\dagger} \widehat{\mathcal{O}} \rangle$, which includes cross-terms between electric and magnetic field components. 

\begin{figure}[htb]
\centering \includegraphics[width=\linewidth]{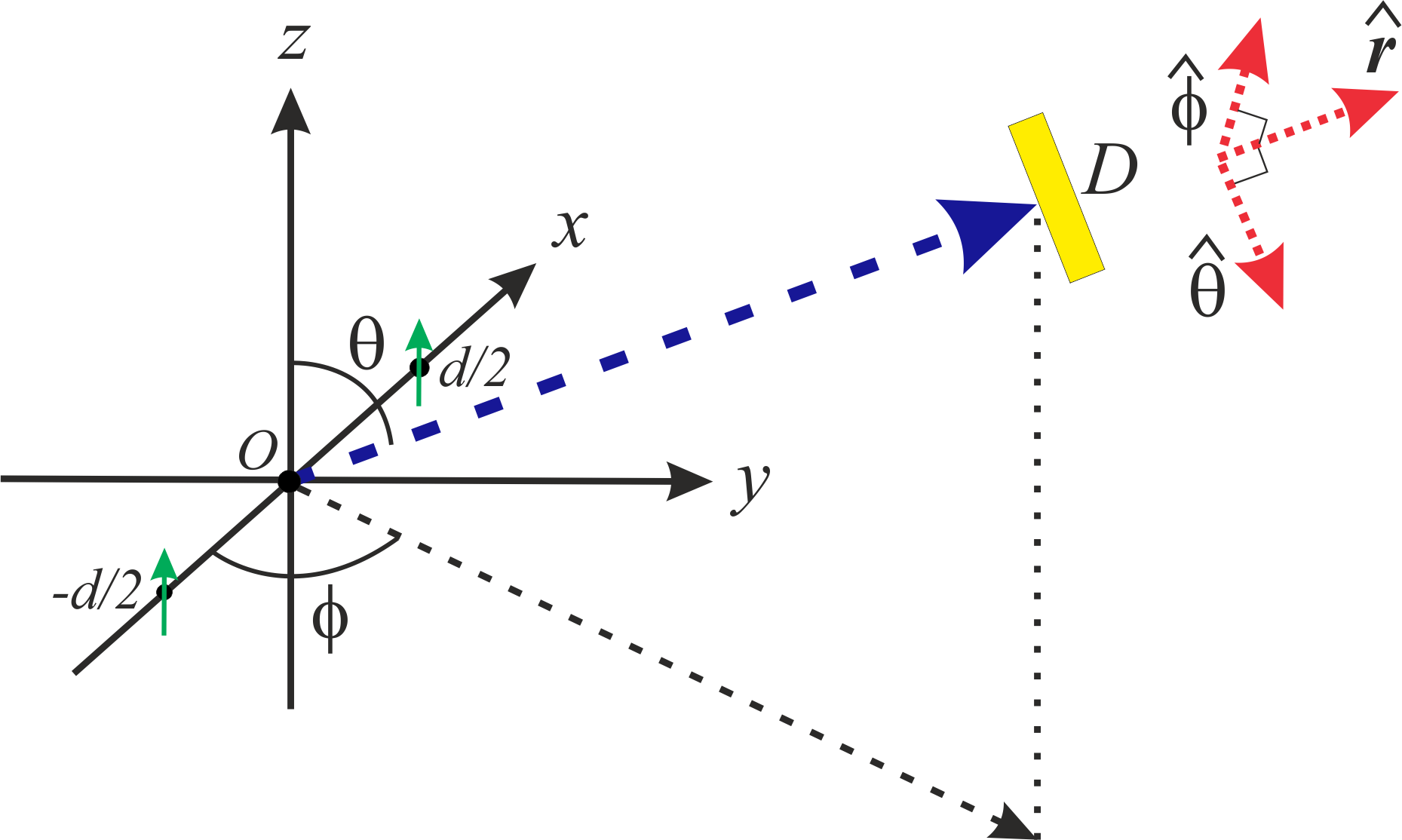} \caption{A pair of in phase Hertzian dipoles, each with electric dipole moments, $\boldsymbol{p} = p\hat{\boldsymbol{z}}  $  at positions $\pm d/2 \hat{\boldsymbol{x}}$, and a single-photon detector, $D$, positioned in a direction, $\theta$, $\phi$ with an orientation perpendicular to the line of sight with the origin $O$. The thin dashed arrow indicates the projection of the segment $[OD]$ on the $xOy$ plane.\label{fig:double_source}} \end{figure}

\textit{Two-dipole model.}~To highlight the consequences of our generalized detection scheme, we study a well-known textbook situation: two in-phase electric dipoles separated by a distance, $d$, along the $x$-axis as illustrated in the schema of Fig.~(\ref{fig:double_source}). Each dipole acts as a Hertzian emitter with dipole moment $\mathbf{p}$ oriented along the $z$-axis. We consider detection of the total radiated field in the far field, using either Glauber's electric-only detector or a more generalized detector that also responds to magnetic fields.

The total electric and magnetic fields in the far-field zone in the direction $\widehat{\mathbf{r}}$ result from a superposition of the two dipole contributions.\cite{Novotny2012} For arbitrary observation directions defined by spherical angles $\theta$ and $\phi$, the far-field electric and magnetic fields take the form:
\begin{align}
\begin{split}
 \boldsymbol{E}(\theta, \phi) &\propto \hat{\boldsymbol{\theta}} \sin\theta \cos\left[ \frac{\delta(\theta, \phi)}{2} \right] \;  \\
 \boldsymbol{B}(\theta, \phi) &\propto \hat{\boldsymbol{\phi}} \sin\theta \cos\left[ \frac{\delta(\theta, \phi)}{2} \right] ,
\end{split} \label{dipmod}
\end{align}
where $\delta(\theta, \phi) = k d \sin\theta \cos\phi$ is the far-field phase difference between the two dipole sources.

For a standard electric-only detector (Glauber model), the detection probability is proportional to $|\boldsymbol{E}|^2 \propto \sin^2\theta \cos^2(\delta/2)$. This vanishes at angles where $\delta = (2n+1)\pi$, corresponding to destructive interference between the two dipole contributions. One must keep in mind however that this result is a consequence of the detector being sensitive only to the electric component of the field.

Applying the generalized detection operator of Eq.~(\ref{Genop}) to the two-dipole radiation fields of Eq.~(\ref{dipmod}), and choosing unit polarization vectors $\mathbf{u}_{e} = \widehat{\boldsymbol{\theta}}$, $\mathbf{u}_{b} = \widehat{\boldsymbol{\phi}}$ appropriate to their respective fields, we find the detection probability:
\begin{align} P_{\rm scat}(\theta, \phi) \propto \sin^2\theta \cos^2\left[\frac{\delta(\theta, \phi)}{2}\right] (1 + |\zeta|^2 + 2\text{Re}[\zeta]) \;, \label{Detprob} 
\end{align}
where the index 'scat' indicates that this result pertains to detection based on radiative scattering. It is of particular interest to remark that for a mixing parameter value of $\zeta = -1$, the detection probability is identically zero for all $\theta$, a phenomenon that is entirely absent from Glauber’s electric-only theory.

This toy model shows that modifying the detector response, specifically by including coherent magnetic dipole interactions, can yield dramatically different detection probabilities, even when the radiated fields are unchanged. Our results underscore the contextual nature of quantum optical measurements and call for further examination of detector-field interactions beyond the standard electric-dipole approximation.

\textit{Absorption and Null Detection Events.}~The \emph{apparently} paradoxical prediction of vanishing detection probability is entwined with subtle issues underlying Glauber's theory and lingering ambiguities in how physicists use the term \emph{absorption}. Glauber-type calculations are often presented as predicting true photon absorption; such as photoelectron emission, joule heating, or other irreversible processes involving the detector’s material degrees of freedom. Inspection of the derivation shows that what is actually computed is the rate of excitation from a ground state to an excited state, similar to the \emph{absorption} coefficient in Einstein's 1916 treatment of black-body radiation\cite{Einstein1916}. The Glauber derivation uses Fermi’s Golden Rule methods to transform the reversible excitation (absorption) into an irreversible process by modeling the final state as being part of a continuum, which allows energy to radiate away from the system. While this procedure effectively introduces irreversibility into the light-matter interaction, it ultimately describes an energy-conserving \emph{scattering} process rather than true absorption. That this methodology succeeds in predicting true absorption rates raises important conceptual, and practical, questions which we now address.

The key lies in the behavior of resonant scattering: in the absence of intrinsic losses, quantum light-matter cross sections reach their unitary limit, with a maximum effective area of $\sigma_{\rm scat} \sim \lambda^2/2$ for a dipole resonance. Electromagnetic scattering calculations show that introducing even a small intrinsic absorption into the material degrees of freedom amplifies its effect dramaticall: resonant re-circulation redirects light that would have been scattered into being absorbed, with minimal impact on power flow into the detector. Consequently, under purely electric-dipole coupling ($\zeta = 0$), the angular and spectral dependence of absorption closely mirrors that of scattering, differing only by a proportionality factor. The extinction (absorption $+$ scattering) cross section, $\sigma_{\text{ext}}$, in a dipole channel is proportional to the imaginary part of the polarizability,
\begin{equation}
\sigma_{\text{ext}} \sim \frac{\omega}{c} \, \operatorname{Im}[\alpha(\omega)] \;,
\end{equation}
and under resonant enhancement scattering and absorption cross sections tend to scale together. Since actual detectors are characterized by their empirical \emph{sensitivity} parameter, $s$, mentioned in Eq.~(\ref{eq:GlauberRule}), the standard Glauber formalism is generally sufficient to model light \emph{absorption} in typical experimental conditions.

This effective equivalence between scattering and absorption breaks down, however, for a value of $\zeta = -1$ where Eq.~(\ref{Detprob}) predicts null detection. This corresponds to perfect destructive interference at the level of the detection amplitude, not at the field level, which is a fundamentally quantum, context-dependent phenomenon. Interpreting this vanishing detection probability to imply the absence of  interaction between the field and the detector would be as misguided as thinking that perfect transmission ($|T|^2 = 1$) in a Fabry-Perot cavity means that the F.P. cavity does not interact with light, the strong internal fields reveal otherwise. In both the perfectly transmitting Fabry-Perot and the perfectly transmitting configuration of the $E-B$ detector described above, complete transparency actually corresponds to maximal interaction, with all incident radiation coherently redirected back into the input channel.\cite{Optinter2016} Introducing even slight absorption into a perfectly transmitting resonant system transforms it from a transparent scatterer into a uniform absorber, eliminating all angular interference signatures, with \emph{no observable interference pattern}.  An ideal, or critically coupled realization could even result in near 100$\%$ quantum absorption efficiency\cite{Grig2015}). A uniform absorption is precisely the distribution that would be predicted were we to replace the two dipole sources with classical particle emitters that radiate \emph{particles} isotropically in the aximuthal angles, but still weighted by a $\sin^{2}\theta$ distribution in the polar angles in accordance with the emission distribution of the individual emitters. This indicates the striking possiblity that the \emph{context} of detection can fundamentally alter the interpretation of how light has \emph{propagated}, shifting its behavior from wave-like to particle-like, purely by the manner in which it is detected.

\begin{figure}[htb] 
\begin{center}

\includegraphics[width=\linewidth]{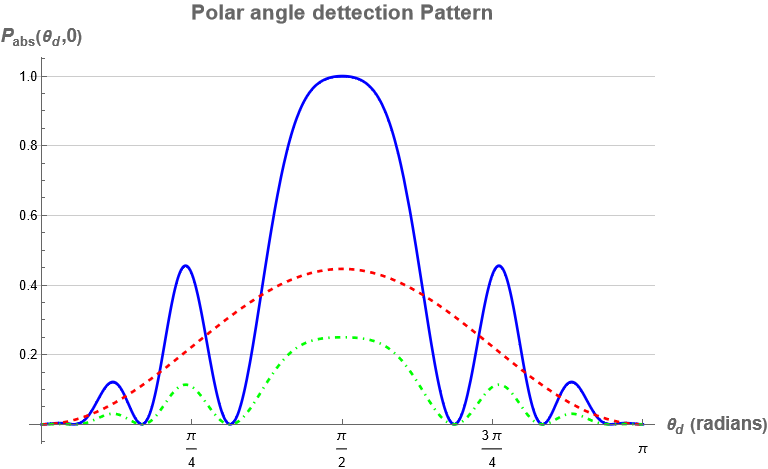}
\vspace{-2mm}
\begin{flushleft}
\small (a)
\end{flushleft}

\vspace{2mm}

\includegraphics[width=\linewidth]{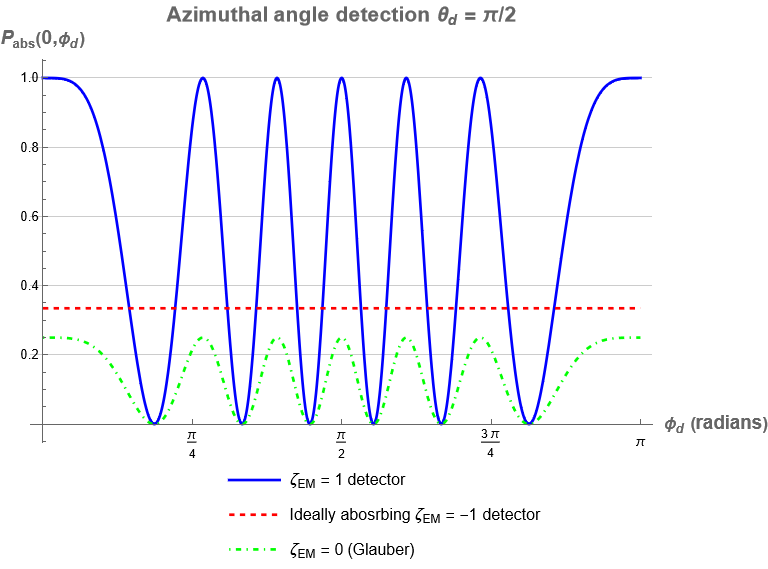}
\vspace{-2mm}
\begin{flushleft}
\small (b)
\end{flushleft}

\caption{Angular detection probabilities in the (a) $xz$ and (b) $xy$ planes, for the Hertzian dipole pair of Fig.~(\ref{fig:double_source}) with $d=3\lambda$. The detection absorption probabilities for the $\zeta=1$ and $\zeta=0$ (Glauber) are plotted as blue (solid), and green (dot-dashed) curves respectively. The surmised absorption curve for a $\zeta=-1$ detector is plotted in red(dashed)}
\label{theta_phi_abs}

\end{center}
\end{figure}

For illustration, the detection probabilities, $P_{\rm abs}(\theta,\phi)$ for the interfering Hertzian dipole model of Fig.~(\ref{fig:double_source}) are plotted in Fig.~(\ref{theta_phi_abs}). For simplicity, we assume that the material losses of the detector are such they achieve their ideal absorption limits where the detector's scattering and absorption cross sections are equal. For $\zeta=1$, the E-B coupling simply increases the detection probability, while for $\zeta=-1$, the prediction is that the interferences in the azimuthal direction can vanish as if corresponding to particle-like emissions with uniform azimuthal emissions. Without specific designs to fix the $\zeta$ parameter, real-world detectors can be expected to exhibit intermediate behaviors. Furthermore, plots of Fig.~(\ref{theta_phi_abs}) are only meant to be indicative of the possiblities offered by meta-material detectors, and more complete calculations measurements are needed in order to assess the extent to which such interference damping effects are observable in practice.

Due to space limitations inherent to a Letter format, our discussion has been primarily heuristic; a more detailed analysis will be pursued in future work. Nonetheless, our conclusions rest on well-established results in quantum and physical optics, and we are confident in asserting the measurement contextuality of light propagation with respect to the detection process.

\textit{Relation to dark-state models.}~In a recent study, Villas-Boas \textit{et al.}~\cite{Villas-Boas2025} propose a picture in which photons propagate in a fundamentally particle-like manner, and interference patterns arise from ‘dark’ and ‘bright’  photonic states. Although our formalisms differ, both formally and conceptually, they share a common feature in that interference effects are produced at the level of light-matter interactions, rather than being viewed as being entirely a property of the field. This shared insight offers new perspectives on long-standing quantum paradoxes such as the delayed-choice quantum eraser.

\textit{Experimental realizations.}~A key prediction of this work is that photodetection may, but not necessarily, exhibit a more ``particle-like'' (interference-suppressed) character \emph{when} the detector couples coherently to both electric and magnetic components of the field. In the optical regime, most conventional detectors rely overwhelmingly on electric dipole transitions. However, realizing the proposed interference control effects requires that both \( \widehat{\boldsymbol{E}} \) and \( \widehat{\boldsymbol{B}} \) contribute meaningfully to the detection amplitude. 

This condition can be met through photonic or plasmonic engineering, as realized in several families of modern detectors. Superconducting nanowire single-photon detectors (SNSPDs), for example, combine ultra-thin absorbing films with back-reflectors and dielectric stacks to maximize $\boldsymbol{E}$-field absorption near the surface, while also supporting circulating supercurrents that effectively couple to the $\boldsymbol{B}$ field \cite{Natarajan2012,Steinhauer2021,Redaelli2024}. Metasurface-enhanced detectors and cavity-integrated architectures can further engineer impedance-matched absorption for both field components \cite{HolzmanIvry2018,MDPI2021}.

Such electromagnetic structuring of detectors has proven critical to surpass the 50\% quantum efficiency ceiling imposed by purely electric, mirror symmetric couplings.\cite{McPhedran1980,Kim2016,Guddala2022} Detectors that combine electric and magnetic sensitivity, either explicitly or via effective impedance engineering, thus offer a promising experimental platform for realizing the context-dependent interference phenomena predicted here. It also appears feasible to carry out demonstrations of this phenomenon in micro-wave analogue experiments where one is able construct macroscopic detectors that are capable of coupling to both electric and magnetic fields. Since photon energies in the microwave regime are small, light is typically treated as a classical wave, the ability to detect light in a particle-like distribution in this domain is an intriguing possibility.

\textit{Conclusion.}~We have presented a generalized quantum model of photodetection that coherently includes both electric and magnetic dipole coupling mechanisms. This extension of the Glauber formalism reveals that interference patterns in quantum optics are not solely determined by the quantum field state, but also by the internal structure and spatial context of the detector itself.

Using a two-source interference model, we showed that the detection probability can be enhanced, or completely suppressed by tuning the relative electric and magnetic sensitivity of the detector, even when the quantum state of the incoming photons is unchanged. This provides a concrete demonstration of measurement contextuality in photodetection, governed by quantum interference between distinct electromagnetic amplitudes.

These results are directly relevant to modern photonic detection technologies, such as superconducting nanowire detectors, cavity-enhanced systems, and metamaterial-integrated platforms, which can be engineered to respond to both electric and magnetic components of light. Our work provides a rigorous framework for interpreting and designing such context-sensitive detection schemes, and suggests new directions for both foundational studies and device optimization in quantum optics.

\begin{acknowledgments}
The author thanks Mohammed Hatifi, Thomas Durt, Nicolas Bonod, Isam Ben Soltane, Remi Colomb, Ross McPhedran, and Jacques Stout, for stimulating discussions. This work was supported by the QuantAMU initiative at Aix-Marseille Université and the Institut Fresnel.
\end{acknowledgments}

\end{document}